\documentclass[trackchanges,twocolumn]{aastex7}

\usepackage{booktabs}
\usepackage{caption}
\usepackage{multirow}
\usepackage{amsmath}
\usepackage{subcaption}
\usepackage{units}
\usepackage{tabularx}
\begin{document}

\title{Kinematic Evidence for Open Cluster Origins of Galactic Binary Neutron Stars}

\author[]{Wen-Jie Yu}
\affiliation{Institute for Frontier in Astronomy and Astrophysics \& Faculty of Arts and Sciences, Beijing Normal University, Zhuhai 519087, China}
\affiliation{School of Physics and Astronomy, Beijing Normal University, Beijing 100875, China}
\email{wenjie.yu@mail.bnu.edu.cn}

\author[0009-0000-1929-7121]{Lu Zhou}
\affiliation{School of Physics and Technology, Wuhan University, Wuhan, Hubei 430072, China}
\affiliation{Institute for Frontier in Astronomy and Astrophysics \& Faculty of Arts and Sciences, Beijing Normal University, Zhuhai 519087, China}
\email{zhoulugz@whu.edu.cn}

\author[0000-0002-3309-415X]{Zhi-Qiang You}
\affiliation{Institute for Gravitational Wave Astronomy, Henan Academy of Sciences, Zhengzhou 450046, Henan, China}
\email{zhiqiang.you@hnas.ac.cn}

\author[0000-0002-9174-638X]{Hao Ding}
\email{hdingastro@hotmail.com}
\affiliation{Korea Astronomy and Space Science Institute, 776 Daedeokdae-ro, Yuseong-gu, Daejeon, Republic of Korea}

\author[0000-0002-0880-3380]{Lu Li}
\affiliation{Shanghai Astronomical Observatory, Chinese Academy of Sciences, 80 Nandan Road, Shanghai 200030, China}
\email{lilu@shao.ac.cn}

\author[0000-0001-8713-0366]{Long Wang}
\affiliation{School of Physics and Astronomy, Sun Yat-sen University, Daxue Road, Zhuhai, 519082, China}
\email{wanglong8@mail.sysu.edu.cn}

\author[0000-0001-7049-6468]{Xing-Jiang Zhu}
\affiliation{Institute for Frontier in Astronomy and Astrophysics \& Faculty of Arts and Sciences, Beijing Normal University, Zhuhai 519087, China}
\email[show]{zhuxj@bnu.edu.cn}


\begin{abstract}
We investigate the potential birthplace of Galactic binary neutron star (BNS) systems through a kinematic analysis. Using high-precision astrometry from Gaia DR3, updated pulsar distances, and Monte Carlo sampling of astrometric errors, we integrate the past trajectories of 11 Galactic BNSs and 167 globular clusters plus 2967 open clusters, to search for past encounters. Our results suggest that BNS origin in globular clusters is unlikely, with low encounter probabilities (e.g., $\lesssim 0.5\%$ for NGC 5139) and requiring excessive ejection velocities. Conversely, our analysis indicates that open clusters are a non-negligible formation channel. Specifically, the double pulsar J0737$-$3039 shows a $13.9\%$ ($5.4\%$) probability of originating from the young cluster OC 0450 (Theia 58). Based on encounter proximity and time, we argue that Theia 58 is its more plausible birthplace. Our work provides kinematic evidence consistent with an open-cluster origin for a subset of field BNSs.
\end{abstract}


\section{Introduction} \label{sec:introduction} 

The discovery of the first BNS system, PSR B1913+16 \citep{hulsetaylor1974}, was a landmark in astrophysics, providing the first indirect evidence for gravitational waves and a cornerstone for testing general relativity. This milestone was dramatically reinforced in 2017 with the first direct detection of a BNS merger, GW170817, by ground-based gravitational wave detectors as well as electromagnetic telescopes on the ground and in space. This event marked the dawn of multi-messenger astronomy with gravitational waves, providing a unique cosmic laboratory to probe extreme gravity, dense matter physics, heavy-element nucleosynthesis, and the engines of short gamma-ray bursts \citep{Abbott2017a,Abbott2017b}.


To fully interpret the rich data from events like GW170817 and to predict populations for future discoveries, a detailed understanding of how BNS systems form is essential. The classical and most studied formation pathway is the isolated evolution of a massive binary star system in the Galactic field. This complex sequence involves finely-tuned phases of mass transfer, common-envelope evolution, and two supernova explosions. The second supernova imparts a natal kick to the newly formed neutron star, which critically shapes the binary's final orbit and systemic velocity if the binary does not get disrupted \citep{evolution3,evolution5,evolution1,evolution2,2017tauris}.

An alternative and well-established channel is dynamical formation in dense stellar environments, particularly globular clusters \citep[GCs;][]{dynamic1,gammaray3}. Here, frequent gravitational interactions can lead to the creation of BNSs through exchange encounters or hardening of primordial binaries. Indeed, two known BNSs reside within GCs: B2127+11C in M15 and J1807$-$2500B in NGC 6544 \citep{2017tauris}. Population synthesis studies further suggest that BNSs with certain properties, such as short orbital periods and high eccentricities, may be formed in GCs before being ejected into the Galactic field \citep{andrews}.

Between the isolated field and the dense cores of GCs lies a third, underexplored environment: open clusters (OCs).
These clusters are natural nurseries for the massive binary star systems that are progenitors of BNSs \citep{Lada03ARAA,KrumJoss19ARAA}. While the high stellar densities necessary for dynamical BNS formation are absent, the primordial binary evolution channel is likely to begin in such stellar groups. As clusters dissolve over tens to hundreds of millions of years, any BNSs formed within them through isolated binary evolution would naturally be released into the Galactic field. This provides a plausible link between the observed field BNS population and a cluster origin, distinct from the rare dynamical channel in GCs. Recent work suggests OCs may contribute non-negligibly to the Galactic pulsar population \citep{zhoulu}, raising the intriguing possibility that they are also viable birthplaces for BNSs.

In this work, we perform the first systematic kinematic investigation into the potential cluster origins—in both GCs and OCs—of Galactic BNS systems. We combine updated BNS distances and proper motions with Gaia DR3 astrometry for 167 GCs and 2,967 OCs. Using Galactic orbit integration and comprehensive Monte Carlo simulations to propagate observational uncertainties, we search for past encounters that would signify a common origin. Our primary goals are to: (1) assess the likelihood of a GC origin for known field BNSs, (2) test the hypothesis that OCs are a significant formation channel, and (3) for promising candidates, derive constraints on the ejection history and kick velocity.

The paper is structured as follows. In Section \ref{sec:data-method}, we describe the observational data for BNSs and star clusters and detail our orbit integration and analysis methodology. Section \ref{sec:Result} presents the results, identifying potential associations with both GCs and OCs. In Section \ref{sec:Discussions}, we focus on the most compelling case, the double pulsar J0737$-$3039, and discuss its formation scenario in detail. We summarize our conclusions and discuss their broader implications in Section \ref{sec:Conclusions}.


\section{Data and Methods} \label{sec:data-method}

We perform a kinematic trace-back analysis to identify past close encounters between BNSs and stellar clusters. This requires three key elements: high-precision astrometric data for BNSs, a complete catalog of star clusters with kinematic parameters, and a robust method to integrate their past orbits within the Galactic potential while accounting for observational uncertainties.

\subsection{Observations of Galactic BNS systems}\label{sec:psr-data}

Currently, around 20 Galactic BNSs are known, two of which reside in GCs \citep[see][and references therein]{ZhuAshton20}. For a meaningful kinematic analysis, complete 6D phase-space information (position, distance, proper motion, and radial velocity) is essential. After excluding systems with insufficient data, we base our study on a sample of 11 field BNSs, whose parameters are listed in Table \ref{tab:bns}.

The primary data source for BNS astrometry is the work of \cite{dinghao}, who provided updated distance estimates. For six BNSs, these distances are derived from precise parallax measurements. For the remaining five, distances are estimated from the dispersion measure using two Galactic free-electron density models \citep{ne2001model,ymw162017model}. The right ascension (RA), declination (Dec), proper motions, and characteristic ages are taken from the Australia Telescope National Facility (ATNF) Pulsar Catalogue\footnote{\url{https://www.atnf.csiro.au/research/pulsar/psrcat}} \citep{atnf}.

A major challenge for BNS kinematic studies is the general lack of direct radial velocity ($v_r$) measurements. We adopt two distinct priors to sample the unknown $v_r$ in our Monte Carlo simulations: (1) a uniform distribution between $-200$ and $200$ km/s, which reflects the broad range of possible pulsar velocities, and (2) the three-component Gaussian distribution Eq.~B.2 of \citet{Ding25}, adopting the ``$N\!=\!3$'' parameters listed in Tab.~B.1 of \citealp{Ding25} as an approximation for each systemic recoil velocity direction of field BNSs and other recycled pulsar systems. 

\begin{table*}[t]
\centering
\setlength{\tabcolsep}{4pt}  
\renewcommand{\arraystretch}{1.2}  
\begin{tabular}{@{} l *{10}{c} @{}}
\toprule
    Name &    RA &  Dec &   pmra &  pmdec      & $d$     & $e$  &$P_{\rm spin}$  &$P_{\rm orb}$   & Age & $v_\perp$ \\
         & (deg) & (deg) & (mas/yr) & (mas/yr) &(kpc) & & (ms)        & (days)  & (Myr) & (km/s) \\  
\midrule
  B1913+16 & 288.867 & 16.1076 &$-0.77^{+0.16}_{-0.06}$  &$0.01^{+0.10}_{-0.17}$ &$4.1^{+2.0}_{-0.7}$  &0.62 &59.0 &0.32 &108 &$141^{+59}_{-43}$\\
  J0737$-$3039A\textbar B &144.448  &$-30.6278$ &$-2.57\pm0.03$ &$2.08\pm0.04$ & $0.74\pm0.06 $ &0.088 &22.7\textbar 2773 &0.10 &204  &$5.9\pm0.5$ \\
  J0509+3801 &77.3833 &38.0217 &$2.9\pm0.1$ &$-5.9\pm0.3$  & $4.2^{+1.6}_{-0.9}$  &0.59 &76.5 &0.38 &153 & $118^{+47}_{-30}$\\
  B1534+12 &234.292  &11.9321 & $1.484\pm0.007$ & $-25.29\pm0.01$ & $0.94^{+0.07}_{-0.06}$ &0.27 &37.9 &0.42 &248 & $102^{+8}_{-7}$\\
  J1930$-$1852 &292.625 &$-18.8628$ &$4.3\pm0.2$  &$-5.2\pm0.4$  & $4.6^{+2.4}_{-1.4}$  &0.40 &185.5 &45.1 &163 & $152^{+91}_{-49}$\\
  J1518+4904 & 229.571  &  49.0762 &$-0.68\pm0.03$  &$-8.53\pm0.04$ & $0.81\pm0.02$ &0.25 &40.9 &8.6  & 23834 &$16.0\pm6$   \\
  \midrule
  J1913+1102 & 288.371  &  11.0349 & $-3.0\pm0.5$  &$-8.7\pm0.1$  &$7.3\pm1.5$ &0.09 &27.3 &0.21 &2687 &$94^{+40}_{-33}$   \\
  J1756$-$2251 & 269.196  & $-22.8665$ & $-2.42\pm0.08$  &$0\pm2$    &$2.6\pm0.6$ &0.18 &28.5 &0.32 &443 &$30^{+16}_{-9}$   \\
  J0453+1559 &  73.4375 &  15.9893 & $-5.5\pm0.5$   &$-6.0\pm4.2$  &$0.6\pm0.4$ &0.11 &45.8 &4.1 &3900 &$26^{+16}_{-12}$   \\
  J1411+2551 & 212.829  &  25.8523 & $-3\pm12$     &$-4\pm9$    &$1.1\pm0.2$ &0.17 &62.5 &2.62 & 10342 &$63^{+47}_{-32}$   \\
  J1829+2456 & 277.389  &  24.896  & $-5.51\pm0.06$  &$-7.8\pm0.1$  & $1.1\pm0.3$ &0.14 &41.0 &1.18 & 12373 &$24^{+7}_{-8}$   \\
\bottomrule
\end{tabular}
\caption{Kinematic and physical parameters of the 11 Galactic BNS systems used in this work. Columns include: Right Ascension (RA), Declination (Dec), proper motion in RA (pmra) and Dec (pmdec), heliocentric distance ($d$), binary orbital eccentricity ($e$), spin period ($P_{\rm spin}$), orbital period ($P_{\rm orb}$), characteristic age of the recycled pulsar, and transverse velocity ($v_\perp$). Distances for the first six BNSs are derived from parallax measurements \citep{dinghao}; for the remaining five, distances are estimated from dispersion measure using either the NE2001 \citep{ne2001model} or YMW16 \citep{ymw162017model} electron density model.}
\label{tab:bns}
\end{table*}

\subsection{Catalogs of Galactic star clusters}\label{sec:cluster-data}

To trace the potential birthplace of BNSs, we require a comprehensive census of stellar clusters within the Milky Way, along with their high-precision kinematic and physical parameters.

To assemble kinematic data for the GC sample, we start from the catalog of 170 Milky Way GCs compiled by \cite{edr3} based on Gaia EDR3. We supplement missing radial velocities by cross-matching with two additional sources: the Gaia DR2-based catalog of \cite{dr2}, containing 154 GCs, and the Gaia DR3-based catalog of \cite{oc}, which includes 132 GCs. Radial velocities are unavailable for eight GCs, resulting in 162 GCs with complete astrometric data. Finally, we add five newly identified GCs from \cite{oc} that are not present in the original 170-cluster list, yielding a final sample of 167 GCs. For clusters where the radius containing 50\% of member stars ($r_{50}$) is not available, we adopt the half-mass radius ($r_{\rm hm}$) from the compilation by \href{https://people.smp.uq.edu.au/HolgerBaumgardt/globular/parameter.html}{Holger Baumgardt}.

The primary source for our OC sample is the extensive Gaia DR3 catalog by \cite{oc}. From their list, we select all clusters with available radial velocity measurements, which are essential for a full 6D phase-space reconstruction. This selection yields a catalog of 2,967 OCs, each with complete astrometry (position, distance, proper motion, radial velocity) and key physical parameters such as age and $r_{50}$.
This is the same catalog used by \citet{zhoulu} in a search for pulsars originating from OCs.

\subsection{Orbit Integration and Encounter Analysis}\label{sec:method} 

We reconstruct the past Galactic trajectories of BNSs and star clusters using the \texttt{Galpy} package\footnote{http://github.com/jobovy/galpy}, integrating orbits within the Milky Way potential. Our approach systematically accounts for observational uncertainties through Monte Carlo sampling.

We adopt the \texttt{MWPotential2014} Galactic potential model in \texttt{Galpy}. The Sun's position is set at a distance of $8.122$ kpc from the Galactic center \citep{sundistance} and $20.8$ pc above the mid-plane \citep{sunheight}. The solar motion in the Galactic rest frame is defined as $[12.9, 245.6, 7.78]$ km/s, with a local circular velocity of 220 km/s. All observed coordinates and proper motions are transformed into Galactocentric coordinates using the Astropy \texttt{SkyCoord} module \citep{Astropy22}.

To incorporate astrometric errors, we perform Monte Carlo simulations. For each BNS/cluster pair, we draw $10^5$ samples from Gaussian distributions defined by their measured parameter values and uncertainties (e.g., distance, proper motion). For each BNS, the unknown radial velocity is sampled from the priors defined in Section \ref{sec:psr-data}.

For each Monte Carlo realization, we integrate the orbits backward in time. The integration for BNS/GC pairs is performed back to the BNS's characteristic age (capped at 1 Gyr for computational stability and reliability), with a time step of $5\times 10^3$ years. For BNS/OC pairs, we integrate back to $T_{\rm end}$, defined as the minimum of the BNS's age (i.e., the characteristic age of the recycled pulsar) and the OC's age, ensuring we do not extrapolate beyond the cluster's lifetime.
We emphasize that, for recycled pulsars, the characteristic age is generally an upper limit on the time since recycling and need not equal the true time since the second supernova of the binary progenitor of a BNS system. In contrast, for OCs the relevant requirement for a physical association is simply that the inferred encounter time is younger than the cluster age, since the cluster must exist at the time of ejection. Accordingly, for BNS/OC pairs we integrate only back to $T_{\rm end}$ to avoid extrapolating beyond the cluster lifetime, and we interpret an OC encounter as a constraint on a recent ejection event rather than as an estimate of the system’s full evolutionary age.

At each time step, we compute the 3D separation, $\Delta (t)$, between the BNS and the cluster center. An encounter is recorded if $\Delta (t) < 3 r_{50}$ (or $3 r_{\rm hm}$). For each BNS/cluster pair, we count the number of Monte Carlo realizations ($N_{\rm enc}$) in which at least one such encounter occurs. The association probability is then calculated as $P = N_{\rm enc} / N_{\rm total}$, where $N_{\rm total}$ is the total number of Monte Carlo realizations. Pairs with an initial probability $P > 0.2\%$ are subject to a second run with $10^6$ realizations to refine the probability and parameter estimates.
The threshold $P > 0.2\%$ is adopted as a pragmatic computational filter: at the initial sampling level, 
$P=0.2\%$ corresponds to 200 encounter realizations, which is sufficient to (i) robustly flag candidates above the Monte Carlo noise floor and (ii) justify a second, higher-resolution run to refine both the probability and the encounter-parameter posteriors. Candidates below this level are typically consistent with rare stochastic hits in a large parameter space and do not materially affect the ranked associations discussed in this Letter.

For each encounter, we store key parameters: the look-back time ($\tau$), the separation ($\Delta$), the relative velocity ($\delta v$), and the implied present-day BNS radial velocity required for the encounter. This ensemble of data allows us to evaluate not just the probability of an encounter, but also its dynamical plausibility (e.g., whether the ejection velocity $\delta v$ is consistent with cluster escape speeds and natal kick constraints).

\begin{table*}[t]
\centering
\setlength{\tabcolsep}{12pt}  
\renewcommand{\arraystretch}{1.3}  
\begin{tabular}{l|c c|c c c c}
\toprule
GC & BNS & Prob & $\tau$ &$\Delta$ &$\delta v$ &$v_{r}$ \\
    &   & ($\%$) & (Myr)  & (pc) & (km/s) & (km/s)\\
\midrule
\multirow{5}{*}{NGC 5139}
&J1913+1102 &0.35 (0.31) &$494.3^{+328.3}_{-405.8}$ &$60.3^{+11.6}_{-19.3}$ &$321.2^{+196.6}_{-91.7}$ &$-61.3^{+79.9}_{-53.4}$\\
&J0737$-$3039 & 0.28 (0.45) &$90.9^{+1.1}_{-1.0}$ &$62.7^{+10.0}_{-18.8}$ &$369.8^{+2.1}_{-1.9}$ &$51.0^{+2.3}_{-2.4}$ \\
&J1829+2456 &0.25 (0.31) &$89.2^{+605.9}_{-3.0}$ &$61.4^{+10.8}_{-18.5}$ &$360.7^{+14.6}_{-8.0}$ &$-33.3^{+6.4}_{-61.7}$\\
&J1411+2551 &0.22 (0.23) &$203.5^{+625.6}_{-109.0}$ &$61.0^{+11.6}_{-19.0}$ &$346.2^{+59.3}_{-37.9}$ &$-4.2^{+56.9}_{-82.6}$\\
&B1534+12  &0.21 (0.12) &$180.2^{+1.3}_{-1.3}$ &$59.0^{+12.9}_{-18.8}$ &$361.3^{+14.7}_{-74.3}$ &$-32.7^{+8.4}_{-143.9}$\\
\bottomrule
\end{tabular}
\caption{Candidate BNS/GC pairs with association probability $> 0.2\%$ based on a uniform radial velocity prior (values in parentheses are for a Gaussian prior). Listed are the GC name, BNS identifier, association probability (Prob), encounter lookback time ($\tau$), separation at the encounter ($\Delta$), relative velocity at encounter ($\delta v$), and the present-day BNS radial velocity ($v_r$) required for the encounter. The last four columns are calculated for the Gaussian radial velocity prior.}
\label{table1:gc_pairs}
\end{table*}

\section{Results} \label{sec:Result}

Using the kinematic trace-back method described in Section \ref{sec:data-method}, we systematically assess the probability of past encounters between 11 Galactic BNSs and 3134 stellar clusters (167 GCs and 2967 OCs). Our results are presented separately for globular and open clusters to highlight the distinct implications of each formation environment.

\subsection{Assessing a Globular Cluster Origin}
\label{sec:results-gc}

After performing $10^5$ Monte Carlo realizations for each BNS/GC pair under the default uniform radial velocity prior, we identify only five associations with probabilities $>0.2\%$, all involving the massive GC NGC 5139 ($\omega$ Centauri). The association probabilities (Prob) between five BNS systems and NGC 5139 range from 0.21\% to 0.35\%, for which the results are summarized in Table \ref{table1:gc_pairs}. Notably, the encounter times ($\tau$) for J0737$-$3039 and B1534+12 are tightly constrained at $90.9^{+1.1}_{-1.0}$ Myr and $180.2 \pm 1.3$ Myr ago, respectively.
The association probabilities for five BNS/GC pairs are comparable while adopting the Gaussian radial velocity prior.

A critical constraint on a dynamical ejection scenario is the relative velocity ($\delta v$) at the time of encounter. For all five BNSs associated with NGC 5139, the derived $\delta v$ is high, ranging from $\sim 320$ to $370$ km/s. These velocities far exceed the cluster's estimated escape velocity ($v_{\rm esc}$ $\sim 62.2$ km/s). This implies that if these BNSs originated in NGC 5139, they must have received extreme natal kicks—likely during the second supernova—or undergone strong dynamical interactions to attain such high ejection speeds. Furthermore, the BNS radial velocities that permit these encounters span a wide range, from approximately $-61$ to $51$ km/s\footnote{In strong three-body interactions, the third body's mass determines the BNS's initial state. Assuming the kinetic energy of the ejected BNS and the third body derives solely from the BNS's released binding energy, a $1~M_\odot$ perturber implies a positive initial binding energy, suggesting BNS formation via three-body capture. Conversely, a massive perturber, such as a $10~M_\odot$ BH, allows for an initially bound BNS binary at the lower limit of $\delta v$.}.

Given the low probabilities ($\leq 0.45\%$) and the physically challenging requirement for ejection velocities several times the cluster's escape speed, we conclude that a GC origin for these field BNSs is unlikely. The detected ``associations" are more consistent with random alignments rather than evidence for a true birth site.

Our conclusion that a GC origin is disfavored for the observed field BNS sample is consistent with previous studies \citep[e.g.,][]{BaeKimMok14,Belczy18bns,YeBNSgc20} finding that the expected contribution of GCs to the overall NS–NS merger population is small compared to isolated binary evolution, even under optimistic assumptions, and that dynamically formed/ejected BNSs from GCs are not expected to dominate the field population.
In our analysis, the kinematic encounter probabilities for GCs are themselves uniformly low; moreover, even if one were to entertain these candidate GC encounters, the implied ejection speeds are several times larger than typical GC escape speeds, providing an independent physical argument against a GC origin for these systems.


\subsection{Evidence for an Open Cluster Origin}
\label{sec:results-oc}

The application of a more physically motivated Gaussian prior for the unknown radial velocities of recycled pulsars \citep{Ding25} provides a robust test for the OC origin hypothesis. Our analysis reveals several significant associations between Galactic BNSs and OCs, with probabilities substantially higher than those found for GCs. The most compelling results are summarized in Table \ref{table3:bns with oc}.
We stress that a kinematic encounter with a present-day OC does not by itself establish a unique birthplace, especially given the large OC sample. Rather, the trace-back analysis identifies candidates that (i) are statistically prominent within the Monte Carlo ensemble and (ii) yield physically plausible encounter parameters. The null test presented in Appendix \ref{app:nulltest} provides an explicit calibration against chance alignments by comparing the baseline probabilities to results from a speed-preserving isotropic control analysis. In this sense, the strongest cases should be viewed as null-calibrated candidates deserving focused physical scrutiny, not as definitive identifications of a single birth cluster.

The double pulsar system J0737$-$3039 emerges as the strongest candidate. It shows a significant association probability of 13.87\% with the young cluster OC 0450, along with a 5.35\% probability with Theia 58. This confirms that the previously identified link \citep{zhoulu} is robust to a more restrictive kinematic prior. J0737$-$3039 also shows non-negligible probabilities ($> 2\%$) with three other OCs (Theia 3397, Theia 3048, and Platais 8), though the association with Platais 8 drops to 1.69\% under the Gaussian prior, making it a less likely candidate due to the high implied radial velocity.

Other BNS systems also show notable connections under the Gaussian prior. J0453+1559 maintains a 4.63\% probability with the extended cluster CWNU 2136 and a 2.75\% probability with the young cluster HSC 1553. J1518+4904 shows a 2.17\% probability of originating from the young, low-mass cluster OC 0279. While the probabilities for B1534+12 with its candidate OCs (COIN-Gaia 9, CWNU 405, UPK 300) decrease under the Gaussian prior (to $\sim$0.7-1.1\%), they remain non-zero. B1913+16 and J0509+3801 each maintain $\sim 2\%$ probability associations with their respective candidate OCs.

The candidate OCs exhibit considerable diversity in their properties. Their masses span from approximately $100 M_{\odot}$ (Platais 8) to nearly $1800 M_{\odot}$ (Theia 3318). Their half-mass radii ($r_{50}$) range from compact (5.7 pc for OC 0279) to extended (44.4 pc for CWNU 2136). This suggests that BNS formation may not be restricted to a specific type of OC but could occur across a range of cluster environments.

The significantly higher probabilities ($\gtrsim 5\%$ for the strongest candidates) for OC associations compared to GCs ($\leq 0.45\%$) under the same analysis framework, combined with physically plausible ejection velocities (explored in Section \ref{sec:Discussions}), provide the first kinematic evidence that OCs are a viable and potentially important birthplace for Galactic BNS systems. The consistency between the uniform and Gaussian prior results for the most promising candidates strengthens the reliability of these associations.

One might ask whether the higher incidence of OC associations relative to GCs is driven purely by the fact that the OC catalog is larger (2967 OCs versus 167 GCs). To quantify this, we perform a control experiment in which the BNS speed distribution is preserved but the velocity direction is randomized (Appendix \ref{app:nulltest}). This null test estimates the expected rate of trajectory–cluster encounters arising from geometric chance given the size of the catalog. We find that the strongest OC candidates for J0737$–$3039 remain enhanced by factors of 
$\sim 10-10^{2}$ relative to this control, indicating that the leading OC associations are not simply a consequence of the larger number of OCs.

\begin{table*}[t]
    \centering
    \setlength{\tabcolsep}{6pt}
    \renewcommand{\arraystretch}{1.1}
    \begin{tabular}{@{} l|cc|cccccc @{}}
        \toprule
        BNS & OC name & Prob & $D$ & $\mu_{\alpha}$ & $\mu_{\delta}$ & Age  & $r_{50}$ & $M_{\rm tot}$ \\
         & & ($\%$) & (kpc) & (mas/yr) & (mas/yr) & (Myr)  & (pc) & ($M_{\odot}$) \\
        \midrule
        \multirow{5}{*}{J0737$-$3039}
&OC~0450 & 12.42 (13.87) & $0.3236 \pm 0.0001$ & $-11.40 \pm 0.02$ & $4.66 \pm 0.02$ & $21.8 $  & 17.78 & 251.32 \\
&Platais~8 & 5.61 (1.69) & $0.1336 \pm 0.0000$ & $-16.24 \pm 0.11$ & $13.74 \pm 0.12$ & $31.2$  & 6.85 & 98.75 \\
&Theia~58 & 3.69 (5.35) & $0.4843 \pm 0.0004$ & $-8.82 \pm 0.02$ & $5.52 \pm 0.02$ & $23.6 $  & 10.36 & 298.09 \\
&Theia~3397 & 2.47 (5.46) & $1.0374 \pm 0.0017$ & $-4.89 \pm 0.01$ & $4.01 \pm 0.01$ & $60.3 $  & 21.15 & 482.84 \\
&Theia~3048 & 2.01 (4.66) & $3.1257 \pm 0.0250$ & $-2.15 \pm 0.01$ & $2.45 \pm 0.01$ & $649 $  & 25.60 & 845.78 \\
        \midrule
    \multirow{3}{*}{J0453$+$1559}
&HSC~1553 & 4.68 (2.75) & $0.2119 \pm 0.0001$ & $5.63 \pm 0.09$ & $-16.33 \pm 0.13$ & $38.5 $  & 9.56 & 232.07 \\
&CWNU~2136 & 2.51 (4.63) & $1.2767 \pm 0.0056$ & $0.73 \pm 0.02$ & $-5.57 \pm 0.02$ & $331$  & 44.40 & 375.45 \\
&CWNU~522 & 2.24 (2.56) & $0.3811 \pm 0.0004$ & $0.88 \pm 0.03$ & $-4.48 \pm 0.03$ & $35.8 $  & 10.34 & 132.17 \\
        \midrule
    \multirow{1}{*}{J1518$+$4904}
&OC~0279 & 4.39 (2.17) & $0.2807 \pm 0.0002$ & $6.30 \pm 0.03$ & $-9.80 \pm 0.03$ & $17.0 $  & 5.68 & 114.85 \\
        \midrule
    \multirow{3}{*}{B1534$+$12}
&COIN-Gaia~9 & 3.95 (1.07) & $0.8686 \pm 0.0015$ & $-2.04 \pm 0.01$ & $-3.07 \pm 0.02$ & $150 $  & 12.94 & 503.28 \\
&CWNU~405 & 2.67 (0.70) & $1.1260 \pm 0.0030$ & $-0.52 \pm 0.01$ & $-3.99 \pm 0.02$ & $73.2$  & 13.02 & 401.81 \\
&UPK~300 & 2.04 (0.65) & $0.9169 \pm 0.0039$ & $0.21 \pm 0.04$ & $-0.77 \pm 0.02$ & $62.4$  & 6.65 & 116.80 \\
        \midrule
    \multirow{1}{*}{B1913$+$16}
&CWNU~1074 & 2.62 (1.95)  & $0.7268 \pm 0.0010$ & $0.36 \pm 0.02$ & $-2.04 \pm 0.01$ & $36.4 $  & 16.34 & 376.34 \\
        \midrule
        \multirow{1}{*}{J0509$+$3801}
&Theia~3318 & 2.44 (2.36)  & $2.9122 \pm 0.0242$ & $-0.65 \pm 0.01$ & $-0.79 \pm 0.01$ & $52.7$  & 22.74 & 1776.52 \\
        \bottomrule
    \end{tabular}
    \caption{Candidate BNS/open cluster (OC) pairs with association probability $> 2\%$ under a uniform radial velocity prior (values in parentheses are for a Gaussian prior). Columns include BNS name, OC name, association probability, OC distance ($D$), proper motion components ($\mu_{\alpha}$ and $\mu_{\delta}$), cluster age, half-mass radius ($r_{50}$), and total mass ($M_{\rm tot}$).}
    \label{table3:bns with oc}
\end{table*}

\begin{table*}[t]
\centering
\setlength{\tabcolsep}{7pt}  
\renewcommand{\arraystretch}{1.2}  
\begin{tabular}{@{} c|*{5}{c} |ccc  @{}}
\toprule
 & \multicolumn{5}{c|}{Estimated ejection scenario} & \multicolumn{3}{c}{Present BNS parameters} \\
\midrule
OC name & $\tau$ & $\Delta$  & $v_{\rm sp,oc}$ &$v_{\rm sp,bns}$ &$\delta v$   &$d$ &$v_{r}$ &$v_{\rm sp}$ \\
       & (Myr)  & (pc) & (km/s) &(km/s) & (km/s)   & (kpc) & (km/s) & (km/s) \\
\midrule
 OC~0450  & $7.7^{+1.5}_{-1.5}$ &$39.1^{+10.1}_{-13.9}$ &$64.4^{+8.6}_{-8.8}$ &$93.4^{+7.8}_{-7.3}$ &$59.4^{+14.6}_{-10.0}$ &$0.74^{+0.06}_{-0.06}$ &$61.5^{+19.5}_{-13.8}$ &$61.7^{+19.4}_{-13.7}$\\

Theia~58  &$7.2^{+0.7}_{-0.7}$ &$23.0^{+5.7}_{-8.0}$ &$76.7^{+4.1}_{-4.0}$ &$87.4^{+7.1}_{-7.3}$ &$45.9^{+7.1}_{-6.1}$ &$0.74^{+0.06}_{-0.06}$ &$56.1^{+9.0}_{-8.1}$ &$56.4^{+8.9}_{-8.1}$\\

Platais~8 &$5.2^{+0.6}_{-0.6}$ &$18.4^{+1.6}_{-2.8}$ &$48.5^{+3.1}_{-3.3}$ &$160.1^{+17.4}_{-11.4}$ &$127.7^{+19.5}_{-12.9}$ &$0.79^{+0.05}_{-0.04}$ &$144.4^{+22.7}_{-14.1}$ &$144.5^{+22.7}_{-14.1}$\\

Theia~3397 &$14.8^{+1.3}_{-1.2}$ &$45.9^{+12.5}_{-18.0}$ &$123.0^{+6.1}_{-5.8}$ &$95.2^{+7.4}_{-7.2}$ &$34.6^{+2.7}_{-2.3}$ &$0.74^{+0.06}_{-0.06}$ &$-2.7^{+5.5}_{-5.4}$ &$7.3^{+3.1}_{-1.2}$\\

Theia~3048 &$26.6^{+1.2}_{-1.6}$ &$57.7^{+13.5}_{-18.8}$ &$225.5^{+3.4}_{-3.5}$ &$141.16^{+6.7}_{-6.8}$ &$88.1^{+6.0}_{-5.1}$ &$0.74^{+0.06}_{-0.06}$ &$-24.3^{+3.8}_{-3.9}$ &$25.0^{+3.8}_{-3.6}$\\

\bottomrule
\end{tabular}
\caption{Ejection parameters for the double pulsar J0737–3039 and its candidate birth clusters, assuming a Gaussian radial velocity prior. Quantities include the encounter lookback time ($\tau$), separation at the encounter ($\Delta$), space velocities of the cluster ($v_{\rm sp,oc}$) and BNS ($v_{\rm sp,bns}$) at encounter, relative ejection velocity ($\delta v$), present BNS distance ($d$), predicted current radial velocity ($v_{r}$), and current space velocity ($v_{\rm sp}$).}
\label{table4:encounter parameters}
\end{table*}

\begin{figure}[htbp]
    \centering
    \includegraphics[width=\linewidth]{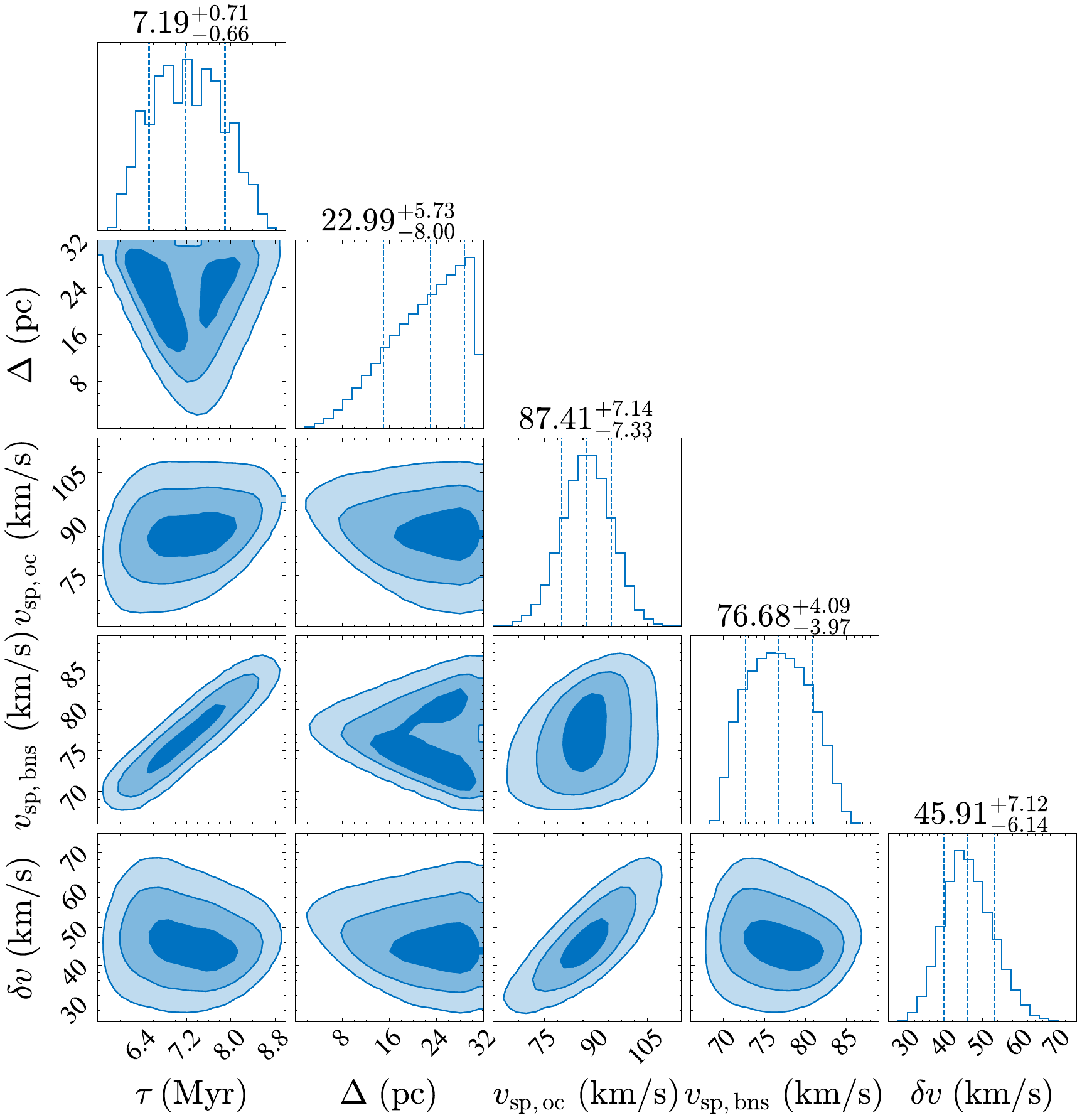}
    \caption{Distributions of encounter lookback time ($\tau$) and ejection parameters for the pair J0737$–$3039 and Theia 58, as listed in Table \ref{table4:encounter parameters}.}
\label{fig1:cornerj0737_theia58}
\end{figure}

\begin{figure*}[!htbp]
    \centering
    \includegraphics[width=0.98\textwidth, trim=0 0 0 0, clip]{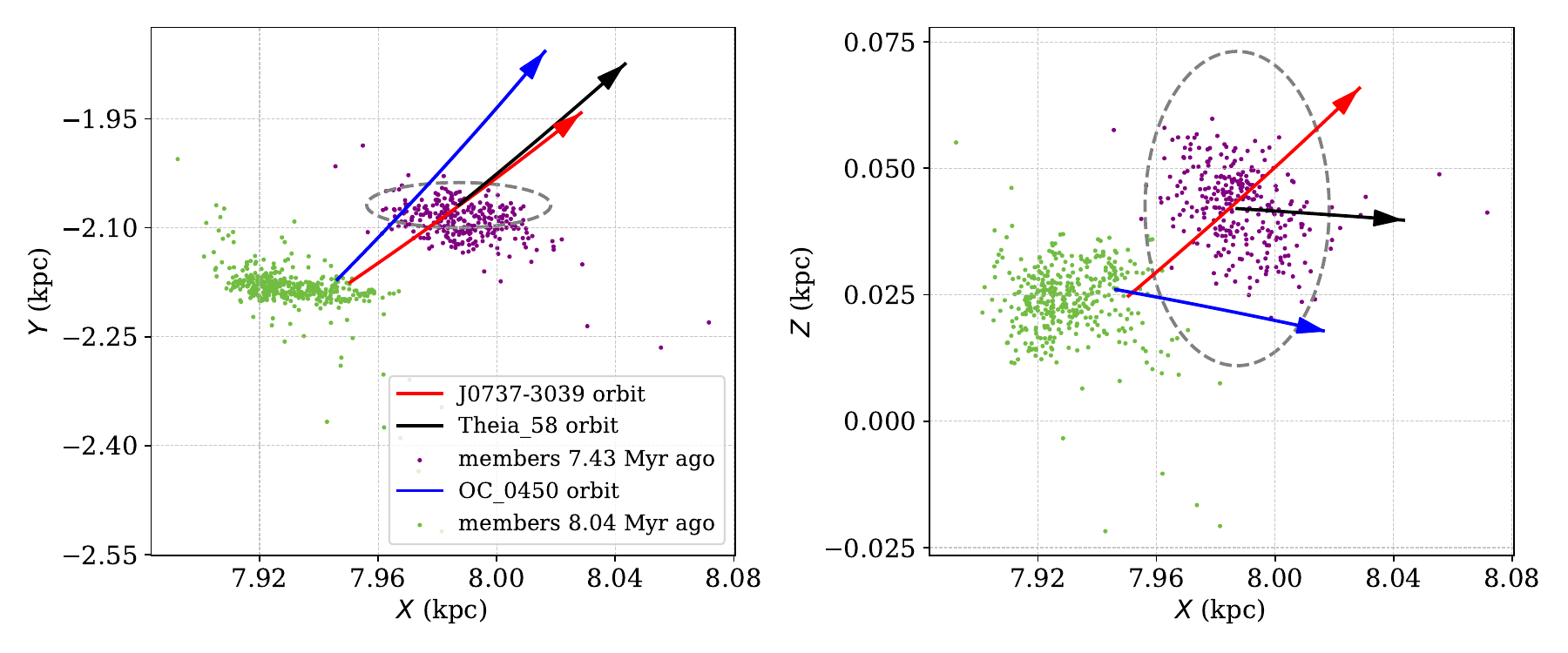}
    \caption{Past trajectories of the double pulsar J0737–3039 (red line) and its two candidate birth open clusters, OC 0450 (blue line and green points for member stars) and Theia~58 (black line and magenta points for member stars). The dashed grey circle marks the $3r_{50}$ boundary of Theia 58. The assumed present BNS radial velocity is 66.25 km/s. The trajectories intersect Theia~58 at 7.43 Myr and OC~0450 at 8.04 Myr lookback time.
}
    \label{fig:orbits_two_OC}
\end{figure*}

\section{The Double Pulsar: A Case Study in Open Cluster Formation} \label{sec:Discussions}

The kinematic analysis presented in Section \ref{sec:Result} demonstrates that OCs might be a significant formation channel for Galactic BNS systems. In this section, we conduct a detailed analysis of the most compelling case—the double pulsar J0737$-$3039—to construct a coherent formation scenario.

The double pulsar J0737$–$3039A/B is a unique astrophysical laboratory. Discovered in 2003–2004 \citep{double_pulsar2003,double_pulsar2004}, its 2.4-hour orbital period and low eccentricity ($e=0.088$) make it one of the most relativistic BNSs known, with an estimated merger time of 86 Myr. Pulsar A is a recycled 22.7-ms pulsar with a characteristic age of 204 Myr, while pulsar B is a slow (2.77-s) pulsar with one of the lowest precisely measured neutron star masses ($1.249 M_{\odot}$) and a characteristic age of 49.3 Myr.
The characteristic ages of the pulsars A/B should not be compared directly to the ages of the candidate birth OCs. The cluster age constrains whether the cluster could have hosted the progenitor at the time of the ejection-triggering event (here, the second supernova), whereas the characteristic age reflects pulsar spin evolution and can differ substantially from the true time since recycling. Our trace-back results therefore test whether the system could have been ejected within the past $\sim 5-30$ Myr from a cluster that existed at that epoch, not whether the BNS “formed” at the cluster’s present age.

Our trace-back analysis identifies five candidate birth clusters for this system. Association probabilities and properties of the cluster are listed in Table \ref{table3:bns with oc}.
Here we discuss the candidate clusters and the BNS formation scenario based on the results derived for a Gaussian radial velocity prior (see Table \ref{table4:encounter parameters}).
First, we disregard Platais 8 as the potential birth cluster because the required radial velocity is high and thus unlikely given our astrophysical prior.
While the remaining four clusters are viable candidates, a fundamental principle guides our interpretation: a binary system can have only one birth site. Therefore, the multiple significant associations likely indicate residual uncertainties in the kinematic trace-back (e.g., from the unknown radial velocity) rather than a true dynamical history involving multiple clusters.
Given this, we focus on the first two encounters, OC 0450 and Theia 58, in the following discussion.
The posterior distributions for the encounter parameters of J0737$–$3039 and Theia 58 are shown in Figure \ref{fig1:cornerj0737_theia58}.
The past trajectories of J0737$–$3039 and its candidate birth clusters are visualized in Figure \ref{fig:orbits_two_OC}.

The encounter parameters for OC~0450 and Theia~58 are physically plausible, and the derived ejection velocities for each ($\delta v \sim 60$ km/s for OC 0450; $\delta v \sim 46$ km/s for Theia 58) fall within the broad range allowed by independent constraints on the natal kick for pulsar B \citep{pulsarb2005,kickconstrain}.
In our analysis for J0737–3039, the closest and most temporally well-defined encounter occurs with Theia 58 at 
$\tau = 7.2\pm0.7$ Myr. The subsequent close approach to OC 0450 ($\tau \sim 7.7$ Myr) occurs slightly later in the backward integration.
Furthermore, the ejection from Theia 58 occurred at a smaller separation ($\sim 23$ pc) and with a lower $\delta v$ of $45.9^{+7.1}_{-6.1}$ km/s.
We therefore consider Theia 58 the more likely birthplace\footnote{We speculate that, given similar age and kinematic properties, it is possible that OC 0450 and Theia 58 share a common origin.} of J0737–3039.

Integrating our kinematic constraints with established binary evolution models, we propose the following formation history for J0737–3039.

\begin{itemize}
    \item Birth in Theia 58: A massive binary system was formed within the young OC Theia 58, approximately 24 Myr ago.

\item Isolated Binary Evolution in situ: The system underwent stable mass transfer and a first supernova, forming pulsar A and recycling it, all while likely still bound to the cluster.

\item Ejection at the Second Supernova: Approximately 7.2 Myr ago, the progenitor of pulsar B—a low-mass helium star as indicated by its low neutron star mass \citep{dewi2004,willems2004}—underwent a supernova. The associated asymmetric natal kick, $\sim 46$ km/s as informed from our analysis, was sufficient to unbind the BNS from Theia 58's potential but not so large as to disrupt the tight binary. This kick imparted the system's current systemic velocity.


\end{itemize}

Finally, assuming the ejection time $\tau$ corresponds to the spin-down age of both pulsars since the second supernova, we can estimate their initial spin periods. Following \citet{ageconstrian}, using a constant braking index model and adopting a flat distribution for the braking indices ($n_{A} \in [0,5]$ and $n_{B} \in [1.4,3]$), we derive the post-supernova spin periods: $P_{0,A} = 22.30\pm 0.04$ ms and $P_{0,B} = 2564\pm 20$ ms.
We note that pulsar B's initial period lies in the upper end of the initial spin period distribution of neutron stars inferred in \citet{Du24initial}.

\section{Conclusions} \label{sec:Conclusions}

In this study, we have conducted the first systematic kinematic investigation into the cluster origins of Galactic BNS systems. By combining high-precision Gaia DR3 astrometry for 3134 stellar clusters with updated distances and kinematics for 11 Galactic BNSs, we performed backward orbit integration within a realistic Galactic potential, employing a Monte Carlo framework to account for observational uncertainties.

Our principal findings are twofold.
First, GCs are an unlikely origin for field BNSs. While some past encounters with massive clusters like NGC 5139 ($\omega$ Cen) were identified, the association probabilities are low ($\lesssim 0.45\%$). More critically, the relative velocities required for ejection from NGC 5139 (321–370 km/s) vastly exceed its escape speed, implying a need for extreme ejection conditions.
Second, OCs are a viable and potentially significant formation channel. In stark contrast, we identified multiple compelling associations between BNSs and OCs, with probabilities often exceeding 5\% for the strongest candidates. This provides kinematic evidence that a non-negligible fraction of Galactic BNSs may have formed within OCs.
More generally, our kinematic associations with present-day young OCs should be interpreted as candidate recent ejection sites. They do not exclude formation in older clusters (including dissolved OCs or GCs) at earlier times, but they do indicate that a subset of field BNSs is consistent with having spent part of its early evolution in a cluster environment before entering the field.

The case of the double pulsar J0737$-$3039 offers a detailed proof of concept. A holistic analysis of ejection dynamics, encounter time, and cluster properties favors its origin in the young cluster Theia 58. We propose a coherent formation history: the system formed via isolated binary evolution within the cluster and was ejected 7.2 Myr ago by a modest natal kick during the electron-capture supernova that formed pulsar B.

This work has several important implications.
First, we establish a third, intermediate pathway for BNS formation—primordial binary evolution in OCs—distinct from both isolated field evolution and dynamical formation in GCs.
Second, it naturally connects the observed field BNS population to the dominant birth environment of massive stars, suggesting many ``field" systems might be cluster ejecta.
Third, the methodology itself provides a new independent tool to constrain natal kick velocities, a critical but poorly understood parameter in compact binary astrophysics.

Future progress will be driven by improved astrometry and larger samples. Upcoming Gaia data releases and next-generation radio VLBI programs will deliver more precise proper motions and parallaxes for both pulsars and clusters, substantially reducing the uncertainties in kinematic trace-back. A growing census of Galactic BNSs from pulsar surveys will provide better statistics to quantify the OC contribution. Ultimately, coupling these observational constraints with detailed binary population synthesis models that include cluster physics will be essential to build a predictive, quantitative framework for the formation of compact binaries across the Galaxy.


\begin{acknowledgments}
This work is supported by the National Key Research and Development Program of China (No. 2023YFC2206704), the National Natural Science Foundation of China (Grant No.~12203004), the Fundamental Research Funds for the Central Universities, and the Supplemental Funds for Major Scientific Research Projects of Beijing Normal University (Zhuhai) under Project ZHPT2025001. We thank the referee for very useful comments. Li, L. thanks the support of NSFC No. 12303026 and the Young Data Scientist Project of the National Astronomical Data Center. You, Z.-Q. is supported by the National Natural Science Foundation of China (Grant No.~12305059); The Startup Research Fund of Henan Academy of Sciences (No.~241841224); The Scientific and Technological Research Project of Henan Academy of Science (No.~20252345003); Joint Fund of Henan Province Science and Technology R\&D Program (No.~235200810111). 

\end{acknowledgments}

\software{astropy \citep{Astropy13, Astropy18, Astropy22}, Galpy \citep{galpy}, Matplotlib \citep{Hunter07}, Numpy \citep{HMv+20}.}

\appendix
\twocolumngrid
\setcounter{figure}{0}
\setcounter{table}{0}
\renewcommand{\thefigure}{\thesection\arabic{figure}}
\renewcommand{\thetable}{\thesection\arabic{table}}

\section{Calibrating Candidate Associations with a Monte Carlo Null Test}
\label{app:nulltest}

Because the Milky Way contains thousands of cataloged OCs, orbit integrations that explore observational uncertainties and an assumed radial-velocity prior can yield close approaches between a binary's reconstructed trajectory and \emph{some} present-day cluster even in the absence of a unique physical link. To quantify (i) how frequently the reconstructed trajectory ensemble intersects the cluster catalog under our baseline assumptions, and (ii) how much of that signal exceeds a conservative ``chance'' background, we perform two complementary diagnostics: (1) the any-cluster encounter probability $P_{\rm any}$, which measures the Monte Carlo encounter fraction with a cluster catalog class; and (2) a null-calibrated enhancement test based on speed-preserving isotropic velocity scrambling.

\subsection{Any-cluster encounter probability $P_{\rm any}$}
\label{app:anyprob}

For each binary--cluster pair, our baseline Monte Carlo orbit integration yields an encounter probability $P_i$ defined as the fraction of realizations that satisfy the encounter criterion (minimum separation threshold within the allowed lookback-time window; see main text). We then define the probability that a given binary encounters \emph{at least one} cluster in a catalog class as
\begin{equation}
P_{\rm any} \;=\; 1 - \prod_i (1 - P_i)
\end{equation}
where the product runs over all clusters in the catalog class. $P_{\rm any}$ should be interpreted as an Monte Carlo encounter fraction: given our assumed present-day phase-space distribution for the BNS (measurement uncertainties and radial-velocity prior), it quantifies the fraction of Monte Carlo realizations that yield at least one encounter with the cataloged cluster population. It is not, by itself, a probability of physical origin in a cluster.

Table~\ref{tab:anyprob} lists $P_{\rm any}$ for the 11 BNS systems considered in this work. In the OC case, $P_{\rm any,OC}$ spans $\sim 1\%$--$31\%$ (median $\sim 9\%$), rather than approaching unity for most systems. This indicates that, under our encounter criterion and current astrometric constraints, the method does not trivially return a high probability of encountering ``some OCs'' for every binary, despite the large OC catalog. For GCs, $P_{\rm any,GC}$ is uniformly sub-percent for all systems in Table~\ref{tab:anyprob}.

\begin{table}[t]
\centering
\begin{tabular}{@{} l|cc @{}}
\toprule
\multicolumn{1}{c|}{Binary (PSR)} & $P_{\rm any,OC}$ (\%) & $P_{\rm any,GC}$ (\%) \\
\midrule
J0737--3039   & 30.9 & 0.52 \\
J0453+1559    & 24.9 & 0.09 \\
J1518+4904    &  6.1 & 0.18 \\
B1534+12      & 11.0 & 0.13 \\
B1913+16      &  8.6 & 0.11 \\
J0509+3801    & 20.9 & 0.05 \\
J1930--1852   &  1.2 & 0.21 \\
J1913+1102    &  1.3 & 0.82 \\
J1829+2456    & 15.9 & 0.69 \\
J1756--2251   &  6.9 & 0.22 \\
J1411+2551    &  7.2 & 0.28 \\
\bottomrule
\end{tabular}
\caption{Probability that a given binary encounters at least one OC or GC in the adopted catalogs, computed as $P_{\rm any}=1-\prod_i(1-P_i)$ over all clusters in the catalog class. These values quantify the Monte Carlo fraction of realizations that yield at least one encounter with the cataloged cluster population under the baseline sampling model, and are not by themselves probabilities of physical origin.}
\label{tab:anyprob}
\end{table}

As an illustrative baseline-versus-null comparison (Section~\ref{app:nullmodel}), for J0737--3039 the isotropic control yields $P_{\rm any,OC}\approx 23.6\%$ and $P_{\rm any,GC}\approx 0.12\%$.

\subsection{Null hypothesis: speed-preserving isotropic velocity scrambling}
\label{app:nullmodel}

While $P_{\rm any}$ quantifies how frequently the reconstructed trajectory ensemble intersects the cluster catalog under the baseline sampling model, it does not by itself indicate whether a particular candidate association is exceptional relative to a chance background. We therefore construct a conservative null hypothesis in which each binary's \emph{speed distribution} is preserved but the \emph{velocity direction} is randomized.

For each Monte Carlo realization we: (1) draw astrometric parameters exactly as in the baseline analysis (proper motions sampled from their measurement uncertainties; systemic radial velocity drawn from the adopted prior), which defines a 3D velocity vector and its speed $v$; (2) replace the velocity direction by an isotropically distributed random unit vector $\hat{\mathbf{n}}$ on the sphere, and set $\mathbf{v}'=v\,\hat{\mathbf{n}}$; (3) integrate the resulting trajectory backward in the same Galactic potential and with the same numerical settings as the baseline run; and (4) compute encounter probabilities $P_{\rm rand}$ using the same encounter criterion. This ``speed-preserving isotropic scrambling'' eliminates directional coherence that could encode a genuine dynamical connection to a specific cluster, while retaining the overall kinematic scale of the system. It therefore provides an estimate of the background Monte Carlo encounter fraction expected from geometric chance given the large number of cataloged clusters.

\subsection{Pairwise excess above the null: enhancement ratio $R$}
\label{app:enhancement}

For each binary--cluster pair we compare the baseline encounter probability, $P_{\rm base}$, to the null probability from the speed-preserving isotropic scrambling experiment, $P_{\rm rand}$, and define the enhancement ratio $R = {P_{\rm base}}/{P_{\rm rand}}$. Values $R \approx 1$ indicate consistency with the isotropic null, whereas $R \gg 1$ indicates a strong non-random enhancement in the baseline calculation.

Table~\ref{tab:enhancement_j0737} summarizes this calibration for J0737--3039. For OCs, we list the five highest-probability candidates from the baseline calculation and report $(P_{\rm base}, P_{\rm rand}, R)$ for each. Theia~58 and OC 0450 show large enhancements ($R\sim 60$--70), whereas Theia~3397 is close to the isotropic expectation ($R \simeq 1.6$).

For GCs, the baseline and null experiments yield largely disjoint sets of low-probability candidates, reflecting both the intrinsically small encounter probabilities and finite Monte Carlo sampling. We therefore list the five most prominent candidates appearing in either experiment (i.e., the union of the top candidates). Only NGC 5139 appears in both experiments; for this object we can compute $R$ directly. For candidates that appear in only one experiment, the corresponding probability in the other experiment is not measured in the current runs, and $R$ is therefore left undefined. Overall, GC candidates remain at sub-percent probabilities and show at most modest enhancement, consistent with the interpretation that GC ``associations'' identified from kinematics alone are generally weak without additional supporting evidence.

We emphasize that $R$ is a statistical calibration against a conservative null hypothesis and does not replace physical plausibility checks. In the main analysis we therefore use the null-calibrated ranking together with encounter kinematics (e.g., relative velocity at encounter and encounter time relative to cluster age) when evaluating candidate birth environments.

\begin{table}[t]
\centering
\renewcommand{\arraystretch}{1.15}
\setlength{\tabcolsep}{4pt}

\begin{tabular*}{0.8 \columnwidth}{@{\extracolsep{\fill}}lccc@{}}
\toprule
Open Cluster & $P_{\rm base}$ (\%) & $P_{\rm rand}$ (\%) & $R$ \\
\midrule
OC 0450    & 13.87 & 0.22 & 63.05 \\
Platais 8  &  1.69 & 0.02 & 84.50 \\
Theia~58    &  5.35 & 0.08 & 66.88 \\
Theia~3397  &  5.46 & 3.34 &  1.63 \\
Theia~3048  &  4.66 & 0.34 & 13.71 \\
\midrule
Globular Cluster & $P_{\rm base}$ (\%) & $P_{\rm rand}$ (\%) & $R$ \\
\midrule
NGC 5139   & 0.279 & 0.03 & 9.30 \\
NGC 6388   & 0.093 & \ldots & \ldots \\
IC 1276    & 0.065 & \ldots & \ldots \\
NGC 5286   & 0.049 & \ldots & \ldots \\
Pal 13     & \ldots & 0.05 & \ldots \\
\bottomrule
\end{tabular*}

\caption{Null-test calibration for J0737--3039 candidate associations. $P_{\rm base}$ is the baseline encounter probability from the Monte Carlo orbit integration used in the main analysis; $P_{\rm rand}$ is the probability under the speed-preserving isotropic velocity scrambling null experiment (Appendix~\ref{app:nullmodel}); and $R\equiv P_{\rm base}/P_{\rm rand}$ quantifies enhancement above the null when both probabilities are available. Entries shown as ``$\ldots$'' indicate that the candidate did not appear in the non-zero encounter-probability list of that experiment (given the finite Monte Carlo sample used for that run), so the corresponding $R$ is undefined.}
\label{tab:enhancement_j0737}
\end{table}

\subsection{Interpretation and scope}
\label{app:interpretation}

The null test supports two conclusions: (1) chance-like encounters are expected at low probability levels in large cluster catalogs, and $P_{\rm any}$ provides a compact measure of the overall Monte Carlo encounter fraction under the adopted sampling model; and (2) a small number of candidates are statistically exceptional relative to the isotropic background ($R\gg 1$) and therefore warrant closer physical scrutiny. In the main analysis we therefore treat kinematic associations as \emph{candidates} and apply additional physical plausibility checks (e.g., encounter time within the cluster lifetime and encounter kinematics) when assessing whether a specific cluster could plausibly be related to the binary's history.

\twocolumngrid 

\bibliography{cite}{}
\bibliographystyle{aasjournal}

\end{document}